\documentclass[reprint,amsmath,amssymb,aps,prb,amssymb,superscriptaddress]{revtex4-1}
\usepackage [latin1]{inputenc}
\usepackage{color}
\usepackage{amsmath}
\usepackage{mathtools}
\usepackage{bm}
\usepackage{amsfonts}
\usepackage{dcolumn}
\usepackage{natbib}
\usepackage{bbold}
\usepackage{soul}
\usepackage[dvipsnames]{xcolor}
\usepackage{amssymb}
\usepackage[breaklinks=true,colorlinks,citecolor=blue,linkcolor=blue,urlcolor=blue]{hyperref}

\usepackage{graphicx}
\def\be{\begin{equation}}
\def\ee{\end{equation}}
\def \bea{\begin{eqnarray}}
\def \eea{\end{eqnarray}}

\begin{document}
\title{Observation of highly anisotropic bulk dispersion and spin-polarized topological surface states in CoTe$_2$}
\author{Atasi Chakraborty}
\thanks{These authors contributed equally to this work.}
\affiliation{Department of Physics, Indian Institute of Technology - Kanpur, Kanpur 208016, India}

\author {Jun Fujii}
\thanks{These authors contributed equally to this work.}

\affiliation{Istituto Officina dei Materiali (IOM)-CNR, Laboratorio TASC, in Area Science Park, S.S.14, Km 163.5, I-34149 Trieste, Italy.}

\author{Chia-Nung Kuo}
\affiliation{Department of Physics, National Cheng Kung University, Tainan 70101, Taiwan}
\affiliation{Taiwan Consortium of Emergent Crystalline Materials, Ministry of Science and Technology, Taipei 10601, Taiwan}

\author{Chin Shan Lue}
\affiliation{Department of Physics, National Cheng Kung University, Tainan 70101, Taiwan}
\affiliation{Taiwan Consortium of Emergent Crystalline Materials, Ministry of Science and Technology, Taipei 10601, Taiwan}

\author{Antonio Politano}
\email{antonio.politano@univaq.it}
\affiliation{Dipartimento di Scienze Fisiche e Chimiche (DSFC), Universit\`a dell'Aquila, Via Vetoio 10, I-67100 L'Aquila, Italy}

\author{Ivana Vobornik}
\email{ivana.vobornik@elettra.trieste.it}
\affiliation{Istituto Officina dei Materiali (IOM)-CNR, Laboratorio TASC, in Area Science Park, S.S.14, Km 163.5, I-34149 Trieste, Italy.} 

\author{Amit Agarwal}
\email{amitag@iitk.ac.in}	
\affiliation{Department of Physics, Indian Institute of Technology - Kanpur, Kanpur 208016, India}


\begin{abstract}
We present CoTe$_2$ as a new type-II Dirac semimetal supporting Lorentz symmetry violating Dirac fermions in the vicinity of the Fermi energy. By combining first principle \textit{ab-initio} calculations with experimental angle-resolved photo-emission spectroscopy results, we show the CoTe$_2$ hosts a pair of type-II Dirac fermions around 90 meV above the Fermi energy. In addition to the bulk Dirac fermions, we find several topological band inversions in bulk CoTe$_2$, which gives rise to a ladder of spin-polarized surface states over a wide range of energies. In contrast to the surface states which typically display  Rashba type in-plane spin splitting, we find that CoTe$_2$ hosts novel out-of-plane spin polarization as well. Our work establishes CoTe$_2$ as a potential candidate for the exploration of Dirac fermiology and applications in spintronic devices, infrared plasmonics, and ultrafast optoelectronics. 
\end{abstract}
 
\maketitle
\begin{figure*}
\centering
\includegraphics[width=1.99\columnwidth]{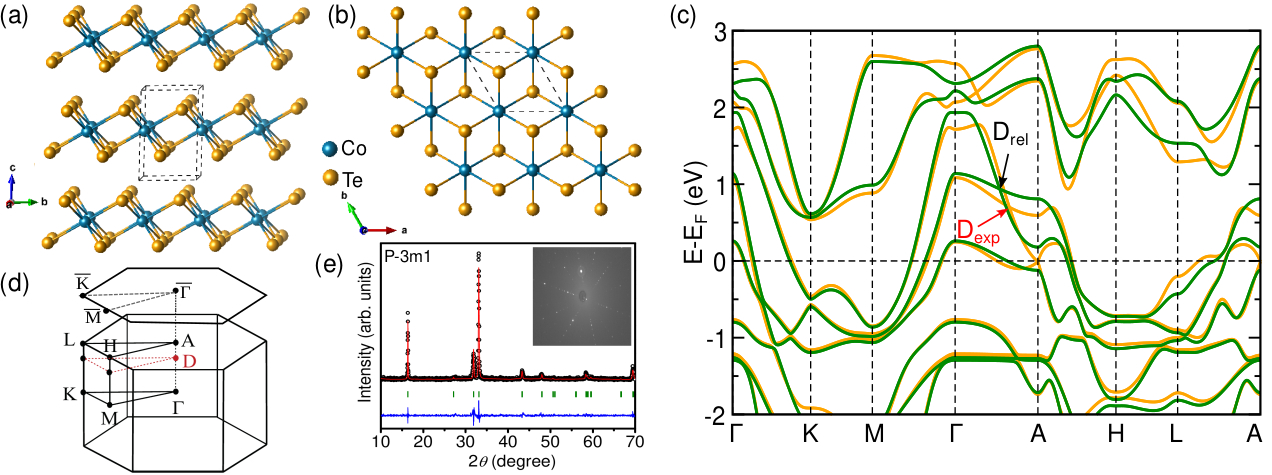}
\caption{\label{fig1} The side (a) and top (b) view of the CoTe$_2$ crystal. In presence of SOC, the band dispersion of the experimental (orange) and relaxed (green) structures are plotted along the high symmetry paths, marked in the  Brillouin zone shown in (d). The type II Dirac crossings near the Fermi energy of the experimental (D$_{exp}$) and relaxed  (D$_{rel}$) structure are marked with red and black arrows. (e) The x-ray diffraction peak structure for CoTe$_2$. The inset shows Laue pattern of the (0001)-oriented CoTe$_2$ single crystals, clearly indicating its purity and the threefold symmetry along the (001) direction.
}
\end{figure*}

\section{Introduction}
The broad class of layered transition metal dichalcogenides (TMDs) has attracted significant attention in the last decades due to 
their novel electronic, optical, and topological properties, combined with their potential for various applications~\cite{2D_TMD_review, TMD_pollution_review, 2D_electronics, TMD_photosensitive_devices, TMD_EEC_review, Vobornik2021K,Olimpio2022E}.  
Owing to the weak inter-layer van der Waals interaction, TMDs offer easy exfoliation of isolated monolayers which host different physical properties from their bulk counterpart. Interesting examples of this include quantum spin Hall effect, superconductivity, charge density wave, and various topological phases~\cite{Tang2017,Chen2018,Liu2012,Miyata2015,Chen2015,Clark_F_2018,Noh_E_2017,Alicea2011,Chiu2021T,Eve2020F,Barun19_NiTe2,Anemone2021E,Yan2017,Zhang2017E}. The physical and chemical properties of TMDs can be tuned by the selection of the constituents, the crystal structures, and the layer thicknesses~\cite{ bandgap_engineering, MoS2_Appl_Mat_today, 2D_TMD_Nature_Rev_Mat, ML_WTe2, ML_MoS2_band_gap,Hlevyack2021D,Mukherjee2020}. 
Specifically, among the TMX$_2$ family of TMDs, PdTe$_2$, PtTe$_2$, PtSe$_2$, and NiTe$_2$ have attracted notable interest due to observation of Lorentz- symmetry violating, type-II Dirac fermions associated with a tilted Dirac cone near the Fermi energy~\cite{Anemone_2020,Huang_T_2016,Fucong2017N,Clark2018F, Antonio2018T,Munisa2021B,Barun19_NiTe2,rizza21_NiTe2,Politano2018D,Xu2018T,Mukherjee2020,Zhang2021H,Casado2022A}. The Lorentz-symmetry breaking type-II Dirac fermions have electronic, optical, and other physical properties which are different from those found in other topological semi-metals. 
The electronic band-structure and spin-polarize topological surface states in these materials have been thoroughly investigated by combining realistic  \textit{ab-initio} calculations with spin-resolved and conventional angle-resolved photoemission spectroscopy (ARPES) experiments.

However, the electronic properties of another prospective candidate material in the series, CoTe$_2$ are yet to be explored \footnote{Zhen Hu1, Libo Zhang, Atasi Chakraborty, Gianluca D'Olimpio, Jun Fujii, Amit Agarwal, Ivana Vobornik, Daniel Farias, Changlong Liu, Chia-Nung Kuo, Chin Shan Lue, Li Han, Kaixuan Zhang, Zhiqingzi Chen, Chenyu Yao, Anping Ge, Yuanchen Zhou, Antonio Politano, Weida Hu, Shao-Wei Wang, Lin Wang, Xiaoshuang Chen and Wei Lu, Terahertz Nonlinear Hall Rectifier Based on Spin-Polarized 1T-CoTe$_2$, To appear in Adv. Mater.}.  CoTe$_2$ can crystallize in both trigonal ($P\bar{3}m1$) and orthorhombic ($Pnn2$ and $Pnnm$) forms. Among these, the centrosymmetric trigonal 1$T$-CoTe$_2$ has recently been shown to be a highly efficient electro-catalyst for water splitting~\cite{CoTe2_catalyst_1, CoTe2_catalyst_2}. In this paper, for the first time, we present a detailed investigation of the electronic structure of 1$T$-CoTe$_2$ by combining first-principles calculations with spin-polarized ARPES experiments. We find that similar to other TMX$_2$ compounds, CoTe$_2$ is also a topological semimetal supporting a type-II Dirac crossing in the vicinity of the Fermi energy. 
%
%
In addition to the bulk electronic structure, we demonstrate that CoTe$_2$ hosts a  ladder of topological surface states arising from several topological band inversions in the bulk electronic structure. These give rise to spin-polarized Dirac surface states, with a large spectral weight. We probe this via spin-ARPES measurements and the measured spin-polarized states 
are consistent with our spin-dependent spectral function calculations. Interestingly, we find that some of the surface states, away from the $\bar{\Gamma}$ point, have an out-of-plane spin polarization.

The rest of the paper is organized as follows. We describe the crystal structure and computational details in Sec.~\ref{S2}, followed by the details of the spin-ARPES measurements in Sec.~\ref{S3}. In Sec.~\ref{S4}, we explore the band structure and geometry of the Fermi surface (FS) in CoTe$_2$. 
We study the origin of the Dirac states, multiple band inversions, and their origin in CoTe$_2$ employing the ARPES measurement combined with \textit{ab initio} electronic structure calculations in Sec.~\ref{S5}. In Sec.~\ref{S6}, we discuss the spin-polarized surface states and the existence of unique out-of-plane spin-polarized states in CoTe$_2$ calculations. We summarize our findings in Sec.~\ref{S7}.

\section{Crystal Structure and Theoretical Methods}
\label{S2}
Bulk CoTe$_2$ crystallizes in  CdI$_2$-type trigonal structure that belongs to the space group $P\bar{3}m1$ (164). Each unit cell of CoTe$_2$ has one Co atom and two Te atoms. To obtain the minimum-energy structure for CoTe$_2$, we performed the symmetry-protected cell volume and ion relaxation using the conjugate-gradient algorithm until the Hellman-Feynman forces on each atom were less than the tolerance value of 10$^{-4}$ eV/\AA~. The cell volume of the experimental structure increased by $2.5 \%$ as a result of the relaxation. The comparison of lattice parameters between experimental and theoretically relaxed structures is presented  in Table~\ref{Tab1}.
\begin{table}
\caption{Comparison of the experimental and theoretically relaxed lattice parameters.}
\label{Tab1}
\begin{tabular}{c c c c} 
 \hline \hline
 &~ {Experimental}~ & ~{Ref}~\footnote{Topological Quantum Chemistry Database}~ &~ {Relaxed}~ \\
 \hline 
 {a/b~(\AA)} & {3.791(9)} & {3.804} & {3.778} \\
 {c~(\AA)}& {5.417(0)} & {5.405} & {5.618} \\
 \hline \hline
\end{tabular}
\end{table}

The trigonally distorted CoTe$_6$ octahedra accommodating the nearest neighbor Co-Te bonds ($\sim 2.55$\AA~) form an edge shared geometrical network on the crystallographic $a-b$ plane [see Fig.~\ref{fig1} (a) and (b)]. Adjacent mono-layers, stacked along the $c$ axis, interact via weak Van-der Waals interaction. Fig.~\ref{fig1} (d) shows the corresponding bulk and (001) surface Brillouin zones (BZs) along with the high-symmetry points.  The CoTe$_2$ crystal structure possesses threefold rotational symmetry around the $z$-axis ($C_{3}$), inversion symmetry $I$, and the three mirror symmetries M$_{100}$, M$_{010}$, and M$_{110}$. Fig.~\ref{fig1} (e) shows the experimental $X$-ray diffraction pattern for CoTe$_2$. The observation of sharp white spots in the Laue diffraction pattern in the inset of Fig.~\ref{fig1} (e) confirms the high quality of the CoTe$_2$ crystals cleaved along the (0001) direction. The presence of the threefold rotation symmetry is also evident.

To perform the {\it ab-initio} calculations, we used the density functional theory (DFT) in the plane wave basis set. We used the Perdew-Burke-Ernzerhof (PBE)~\cite{Perdew1996} implementation of the generalized gradient approximation (GGA)  for the exchange-correlation. This was combined with the projector augmented wave potentials~\cite{paw94, paw99} as implemented in the Vienna {\it ab initio} simulation package (VASP)~\cite{kresse1996efficient, kresse1999ultrasoft}. GGA calculations are carried out with and without Coulomb correlation (Hubbard U) and spin-orbit coupling (SOC). The SOC is included in the calculations as a second variational form to the original Hamiltonian. The kinetic energy cutoff of the plane wave basis for the DFT calculations was chosen to be 450 eV. A  $\Gamma$-centered $12\times12\times 8$  Monkhorst-Pack~\cite{PhysRevB.13.5188} $k$-point grid was used to perform the momentum-space calculations for the  Brillouin zone (BZ) integration of bulk. To calculate the surface spectral function for finite geometry slabs of CoTe$_2$, we construct the tight-binding model Hamiltonian by deploying atom-centered Wannier functions within the  VASP2WANNIER90 \cite{wann90} codes. Utilizing the obtained tight-binding model, we calculate the surface spectral function using the iterative Green's function method, as implemented in the WannierTools package \cite{wanntools}. 

\begin{figure*}[t]
\centering
\includegraphics[width=1.99\columnwidth]{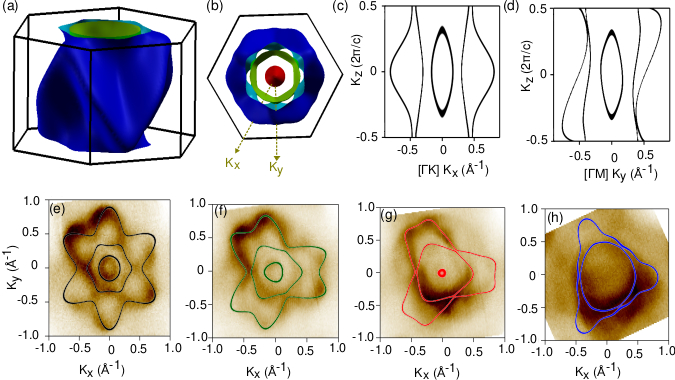}
\caption{\label{fig2} Side (a) and top (b) view of the 3D FS. The projected FS at $E=E_F$ on the (c) $K_x-K_z$ plane along $\Gamma-K$, and (d) the $K_y-K_z$ plane along $\Gamma-M$ directions. The experimentally measured 2D energy contours over $K_x-K_y$ plane at fixed values (e) $K_z=$0.03$ c^*$, (f) 0.16 $c^*$, (g) 0.29$c^*$, and (h) 0.42 $c^*$, where $c^*= 2\pi/c$. The theoretical FS cuts for specific $K_z$ planes are plotted on top of the corresponding experimental results. 
}
\end{figure*}

\section{ARPES and Spin-ARPES measurements} \label{S3}
ARPES and Spin-ARPES measurements were performed at low energy (LE) branch of the APE-NFFA beamline~\cite{Panaccione} at the Elettra synchrotron facility (Trieste, Italy), which is equipped with VESPA ~\cite{Bigi} as an electron spin polarization detector. The details of the experimental geometry, like the electron analyzer slit opening and incoming photon direction with respect to the analyzer lens axis, can be found in Ref.~[\onlinecite{Bigi}]. To determine the inner potential ($V_0$) of CoTe$_2$ (0001) experimentally, angle-resolved valence band spectra and FS maps were measured for the photon energy range between 13 eV and 85 eV with 2 linear polarizations ($s$- and $p$- polarization). Spin-ARPES maps were acquired for two-photon energies ($h\nu$= 19 eV and 75 eV). The energy and angular resolutions for the Spin-ARPES measurements were set to 100 meV and 1.5$^\circ$, respectively. The clean (0001) surface of CoTe$_2$ was obtained by the cleavage of the single crystal \textit {in situ} in an ultra-high vacuum. The sample temperature during the ARPES and Spin-ARPES measurements was kept at 78 K.

\section{Electronic Band-structure and the FS Geometry}
\label{S4}
%
The ionic balance of the chemical formula of CoTe$_2$, suggests that the Co and Te atoms are in \textit{3d$^3$4s$^0$} and \textit{5s$^2$5p$^6$} configurations, respectively. As a consequence, we expect the Co-$d$ and Te-$p$ orbitals to have a major contribution at the Fermi energy ($E_F$). We present the bulk band-dispersion in presence of SOC, for the experimental structure, and also for the relaxed structure in Fig.~\ref{fig1}(c). 
The experimental electronic band dispersion in Fig.~\ref{fig1}(c), clearly shows the existence of a couple of tilted Dirac-like crossings just above $E_F$, along the $\Gamma$-A high symmetry direction. We find that the position of the Dirac point (DP) is sensitive to small variations of the structural parameters. It shifts from $\sim$ 0.68~eV to $\sim$0.92~eV above $E_F$ due to the small change in the structural parameters on relaxation. Since the $\Gamma$-A path is an invariant subspace of the three-fold rotational crystal symmetry ($C_{3}$), the Dirac cone is protected by the rotational symmetry. This is similar to the Dirac crossing in NiTe$_2$ and other related materials in the same space group \cite{Barun19_NiTe2, Nappini20_NiTe2,rizza21_NiTe2}. Two accidental linear band crossings, one exactly at $E_F$ on the high symmetry $A$  point and another one at  $\sim$2.2~eV above $E_F$ along the $\Gamma-A$ path become gapped due to relaxation, as highlighted in Fig.~\ref{fig1}(c). 

The geometry of the FS and its evolution with change in the Fermi energy is shown in Fig.~\ref{fig2} for the relaxed structure. 
The 3D FS for $E=E_F$ is shown  in Fig.~\ref{fig2}(a). The projection of the FS on a plane perpendicular to the $K_z$ axis in Fig.~\ref{fig2}(b) clearly shows three distinct types of band contributions at the FS, each having two-fold degeneracy. Figures \ref{fig2}(c) and (d) capture the projection of the FS on the $K_x-K_z$ plane along the $\Gamma-K$ line, and the $K_y-K_z$ plane along the $\Gamma-M$ direction. 
\begin{figure*}[t]
\centering
\includegraphics[width=1.9\columnwidth]{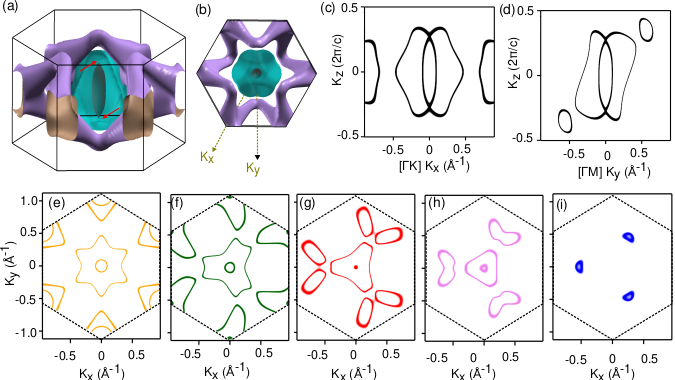}
\caption{\label{fig3} Side (a) and top (b) view of the 3D FS distribution. The planar projection of the constant energy surface at $E-E_F= 0.94$~eV, on the (c) $K_x-K_z$, and the (d) $K_y-K_z$ surface. The 2D energy contours within the Wigner Seitz cell (marked by dotted line) in $K_x-K_y$ plane at fixed $K_z$ values of (e) $K_z=0$, (f) $K_z=0.15~c^*$, (g) $K_z=0.25~c^*$, (h) $K_z=0.30~c^*$, and (i)  $K_z=0.42~c^*$, where $c^*= 2\pi/c$. The 2D plane of (g), which hosts the Dirac point, is marked in Fig.~\ref{fig1} (d).}
\end{figure*}
 The FS along the $\Gamma-M$ path is highly anisotropic as seen in Fig.~\ref{fig2}(d). Clearly, bulk CoTe$_2$ has a strong momentum-dependent anisotropic FS (see Appendix~\ref{appB} for details), which is also expected from the presence of type-II Dirac fermions in the system. To investigate the FS modulations along the $K_z$ direction, we have shown the energy contours at different $K_z$ values in panels (e)-(h) of Fig.~\ref{fig2}. The theoretically calculated (solid line) and the experimentally measured (mud color scale) 2D energy contours within the Wigner Seitz cell are shown in Fig.~\ref{fig2}(e)-(h) over the $K_x-K_y$ plane for different $K_z$ values. Different $K_z$ values are probed in the ARPES  experiment by changing the energy of the incident photon beam. Using the free electron final state model \cite{Damascelli_2004}, we have 
\begin{equation}
    {k}_\perp =\frac{1}{\hbar} \sqrt{2m~(V_0 + E_{\rm kin}\cos^2\theta)}.
\end{equation}
Here, $V_0$ is the inner potential, $E_{\rm kin}$ is the kinetic energy of a photoelectron and $\theta$ denotes the emission angled from the sample surface normal. For the different panels of Fig.~\ref{fig2}(e)-(h), we have $h\nu$ (corresponding $K_z$) = 75 eV (0.03 $c^*$), 70 eV (0.16 $c^*$), 65 eV (0.29 $c^*$), and 60 eV (0.42 $c^*$), respectively where $c^*= 2\pi/c$. We have applied $V_0$ = 11 eV, to calculate the $K_z$ values. 

The experimental FS demonstrates the transformation of its symmetry from sixfold at $K_z=0$ to threefold for $K_z>0$, which is consistent with the theoretical calculations. For $K_z = 0$ [Fig .~\ref{fig2}(e)], all three (the hexapetalus flower-shaped, hexagonal, and circular) states are observed and well matched to the calculated FS. The hexapetalus flower-shaped states in Fig.~\ref{fig2}(e) is transformed into the trefoil in Fig.~\ref{fig2}(f). Due to the experimental geometry and the corresponding matrix-element effect, the measured FS shows an anisotropic distribution in the photoemission intensity. The photoemission intensity is higher along one of the three $\bar{M}$-$\bar{\Gamma}$-$\bar{M}$ directions 
and lower along the two other $\bar{M}$-$\bar{\Gamma}$-$\bar{M}$ directions. This effect reduces the clarity of the three-fold symmetry in the FS, measured for Fig.~\ref{fig2}(g) and (h). However, the strong modulation of the FS on changing $K_z$ is clear, and it is broadly consistent with the 3D FS distribution of Fig.~\ref{fig2}(a). 


\begin{figure*}[t!]
\centering
\includegraphics[width=1.99\columnwidth]{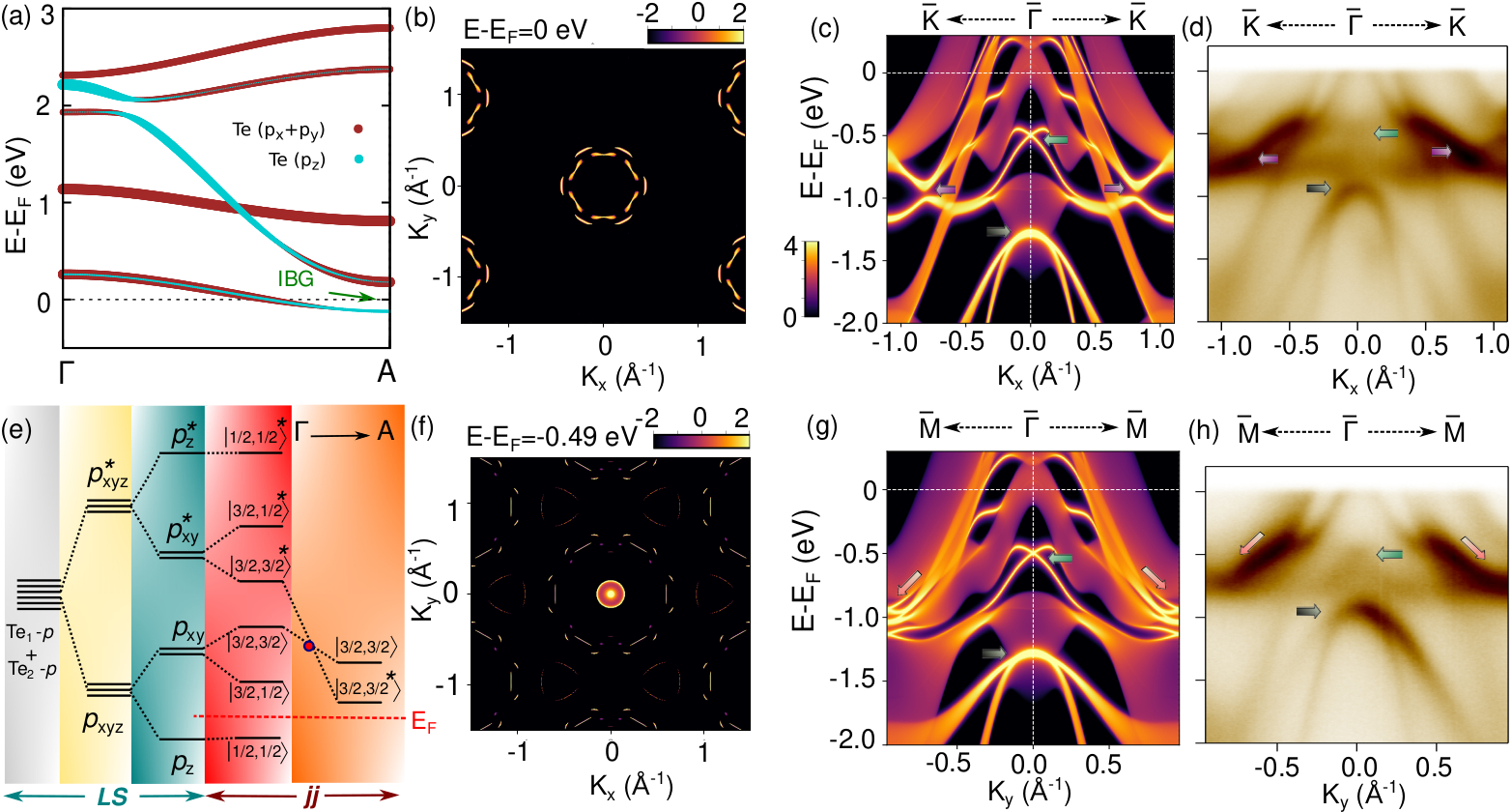}
\caption{\label{fig4} (a) Multiple band inversion arising from the Te-$p$ orbital manifold along the $\Gamma-A$ direction. The band inversion gap near the Fermi energy (IBG) is marked with an arrow. The Fermi arc states at constant energy (b) $E-E_F=0$, and (f) at  $E-E_F=-0.49$~eV. The theoretically calculated momentum resolved spectral density plot along the (c) $\bar{\Gamma}$-$\bar{K}$ and the (g) $\bar{\Gamma}$-$\bar{M}$ paths. The experimentally measured ARPES plots along the (d) $\bar{\Gamma}$-$\bar{K}$ and (h)  $\bar{\Gamma}$-$\bar{M}$ paths. (e) To highlight the origin of the bulk type-II Dirac fermions, we show the schematic of the level diagram of the Te-5$p$ orbitals in presence of a local crystal field and SOC.
}
\end{figure*}

 We now focus on the FS, in the vicinity of the DP. The side and top view of the 3D FS distribution within the Wigner-Seitz cell at $E=E_{DP}$ for the relaxed structure is presented in Fig.~\ref{fig3}(a) and (b), respectively. 
 The presence of three contributing bands, each having twofold Kramer's degeneracy, can be clearly seen. The outer-most part of the FS arises from the electron pocket of the first unoccupied band of CoTe$_2$, as seen in Fig.~\ref{fig1}(c). 
 The type-II nature of the DSM phase is also confirmed by the fact that the Dirac point appears at the four-fold degenerate touching point of the other electron and hole pockets in the middle, as marked by the red arrow in the FS in Fig.~\ref{fig3}(a). 
 The energy contours over the $K_x-K_z$ and the $K_y-K_z$ planes, for $E=E_{DP}$ is shown in Fig.~\ref{fig3}(c) and (d),  respectively. Our calculations reveal a prominent Dirac crossing located at  $K_z\sim\pm 0.25~c^*$. 
 The anisotropic nature of the FS along the $K_y$ direction persists even at the DP. The in-plane projection of the energy contours at the DP is presented in Fig.~\ref{fig3}(e)-(i), for five different out-of-plane distances (or $K_z$ values). 
 At $K_z=0~c^*$, we observe a hexapetalus flower shape along with a small circle at its center. The electron pockets at $K_z=0$ transform into an isolated bean-shaped pattern with increasing $K_z$ magnitude as seen in Fig.~\ref{fig3}(g)-(i). 
 At $K_z$ approaching the vicinity of bulk DP, the central contour converges to a tiny circle while the hexagonal outer contour acquires an almost triangular shape [see Fig.~\ref{fig3}(g) and (h)]. 
 Finally, at $K_z$ = 0.3$c^*$, the FS cut appears as two contours centered around the origin, which exhibit a circular and triangular shape for the inner and outer contours, respectively. 
 In addition, there are small pockets along three of the six $A-H$ lines [see Fig.~\ref{fig3}(h)]. The energy contour at negative $K_z$ values with the same magnitude shows the rest of the three small pockets along the other $A-H$ lines. 
 In Fig.~\ref{fig3}(i) the inner contours vanish and we only see three small pockets along to $A-H$ direction. 

\section{Origin of Dirac states, band inversion, and the Surface States}
\label{S5}
 The presence of type-II Dirac fermions in the bulk dispersion of CoTe$_2$ suggests the strong possibility of finding topologically protected surface states near the Fermi energy. Additionally, the bulk bands of CoTe$_2$ also support several other topological 
 band inversions in its bulk. 
 In Fig.~\ref{fig4}(a), the orbital-resolved band structure along the $\Gamma$-A path shows that the linear crossings near $E_F$ are mainly composed of the Te-5$p$ orbitals. The Dirac band crossing near 0.92~eV above Fermi energy arises from the interplay of the Te $p_x+p_y$ and the Te $p_z$ orbitals. Additionally, we find that these orbitals also contribute to multiple band inversion gaps along different high symmetry paths including $\Gamma-A$ [see Fig.~\ref{fig4}(a)]. To understand the origin of the Dirac band crossing, we show the systematic evolution of the energy levels of the Te-$5p$ orbital manifold in  Fig.~\ref{fig4}(e). The six degenerate $p$ orbital splits into lower (upper) lying three-fold bonding (anti-bonding) orbitals due to inter-site hybridization. The presence of local trigonal distortion of the Co-Te octahedra further lifts the degeneracy of the bonding/anti-bonding $p$ orbitals breaking it into singly degenerate $a_{1g}$ ($p_z$) and doubly degenerate $e_g^\pi$ ($p_x, p_y$) orbitals. Including the SOC splits the $p$ orbitals into fourfold $J_{\rm eff}=3/2$ and two-fold $J_{\rm eff}=1/2$ pseudo spin basis as shown in the fourth column of Fig.~\ref{fig4}(e). The last column of Fig.~\ref{fig4}(e) highlights the effect of the dispersion along the $\Gamma-A$ direction. The bulk type-II Dirac point arises from the crossing of the bonding and anti-bonding states of the $J_{\rm eff}=3/2$ orbitals. 

 The ladder of multiple band inversions and the Dirac point in the bulk band structure points to the existence of topologically protected surface states in CoTe$_2$. This is confirmed by our experiments and theoretical calculations. The measured 
 ARPES results and the corresponding theoretical spectral function of the relaxed structure are shown along the high symmetry $\bar{K}$-$\bar{\Gamma}$-$\bar{K}$ and $\bar{M}$-$\bar{\Gamma}$-$\bar{M}$ directions in Fig.~\ref{fig4}(c), (d) and Fig.~\ref{fig4}(g), (h), respectively. The pattern of the spectral function and position of the surface Dirac cone matches well between the theoretical calculations and experimental plots. However, the other sharp spectral functions [purple, and yellow arrows in Fig.~\ref{fig4}(c), (d) and Fig.~\ref{fig4}(g)] arising from the bulk and surface states are slightly off 
 in energy (see Appendix~\ref{appA} for detailed discussions). This can be due to several reasons including i) small variations in the structural parameters, ii) some ambiguity in the pseudopotential for capturing core states, iii) some impurities or stacking faults in the crystal which are not included in theoretical calculations, amongst others. We also note that as the Bulk Dirac cone is significantly above the Fermi energy, it cannot be directly observed or mapped via our occupied state ARPES data.

The ARPES measurements were done with $h\nu$ = 75 eV, which corresponds to $K_z \sim 0c^*$. Therefore these spectra capture the bulk bands along with the surface states.  
The prominent features corresponding to the surface states, in the measured ARPES spectrum and the calculated spectral function are marked by thick arrows. Despite some discrepancies in the binding energy of a few states, the experimental and the theoretical results show good qualitative agreement. The small energy difference in the location of the surface states possibly arises due to structural effects or from the surface potential which is not included in our theoretical calculations. 

The Dirac cone in the surface states is located at the $\bar{\Gamma}$ point at an energy 0.49~eV below the Fermi energy. The presence of a topological band inversion near $E_F$, as marked by an arrow in Fig.~\ref{fig4} (a) gives rise to this surface Dirac crossing observed in ARPES. 
A similar surface Dirac cone, which has relatively broad features in ARPES experiments compared to theoretical calculations, has also been observed in other isostructural compounds such as NiTe$_2$ and in PtTe$_2$. 
Other than the Dirac cone at the $\bar{\Gamma}$ point,  several sharp non-trivial surface states appear near the high symmetry $\bar{M}$ point and along the $\bar{\Gamma}-\bar{K}$ path. These arise from the multiple band inversions throughout the BZ. 
We find the surface states to be symmetric along both the  $\bar{K}$-$\bar{\Gamma}$-$\bar{K}$ and the $\bar{M}$-$\bar{\Gamma}$-$\bar{M}$ directions. 

The Fermi arc states at constant energy are plotted in Fig.~\ref{fig4} (b) at $E-E_F=0$~eV and in Fig.~\ref{fig4} (f) at $E-E_F=-0.49$ eV. At the Fermi energy, circular arcs of the sharp surface states appear around the $\bar{\Gamma}$ point. In contrast, a prominent peak is observed exactly at the $\bar{\Gamma}$ point Fig.~\ref{fig4} (f)  which captures the dominant surface Dirac, crossing along with a few less intense circular arcs along the  $\bar{\Gamma}$-$\bar{K}$ paths. There is another set of high-intensity surface arc states around $-1.4$ eV below $E_F$ [see Fig.~\ref{fig4} (c) and (g)], which disperse symmetrically around the $\bar{\Gamma}$ point. 
\begin{figure*}
\centering
\includegraphics[width=\linewidth]{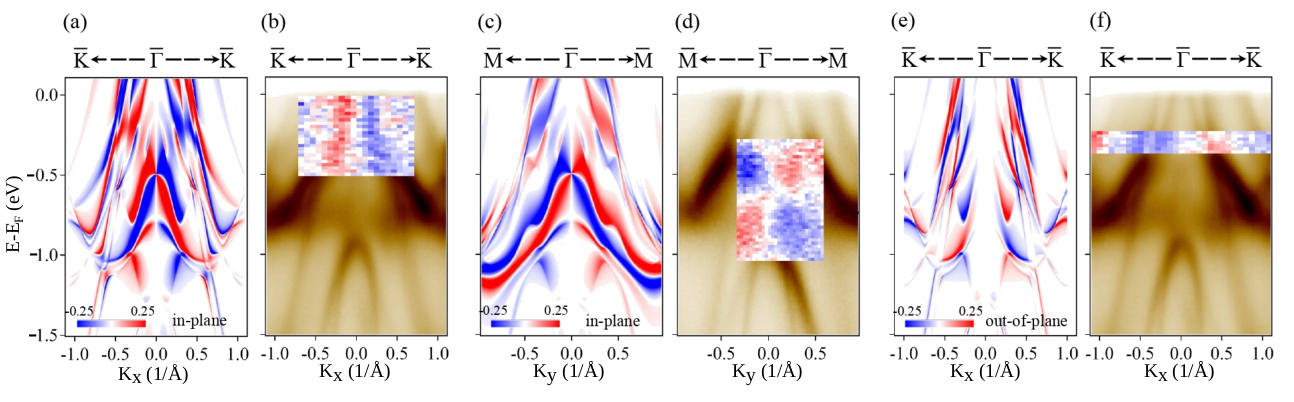}
\caption{\label{fig5}Spin polarization of the surface bands along the 
high symmetry (a) $\bar{K}$-$\bar{\Gamma}$-$\bar{K}$ and (c)  $\bar{M}$-$\bar{\Gamma}$-$\bar{M}$ directions. The spin-ARPES measurements  for (b) $\bar{K}$-$\bar{\Gamma}$-$\bar{K}$ and the (d)  $\bar{M}$-$\bar{\Gamma}$-$\bar{M}$ directions. The spin components are orthogonal to the corresponding momentum directions.
(e) The theoretical and (f) experimentally measured out-of-plane spin polarization along the $\bar{K}$-$\bar{\Gamma}$-$\bar{K}$ direction. CoTe$_2$ supports spin-polarized surface states over a wide range of energies in the entire BZ.}
\end{figure*}
\section{Spin polarized surface states}
\label{S6}
The demonstration of topological surface states in CoTe$_2$ inspires the exploration of their spin-polarization. 
To study the spin polarization of the surface states and the surface Dirac cone, we experimentally measured the spin-polarized ARPES spectrum of CoTe$_2$, as shown in Fig.~\ref{fig5}(b), (d) and (f). The component of probed spin components is chosen to be perpendicular to the direction of the dispersion. The corresponding theoretically calculated spin resolved spectral function is shown in Fig.~\ref{fig5}(a), (c), and (e). 
%
Figures~\ref{fig5}(b) and (d) display the measured spin-resolved band structures superimposed onto the measured spin-integrated band structures shown in Figs.~\ref{fig4}(d) and (h), along $\bar{K}$-$\bar{\Gamma}$-$\bar{K}$ and $\bar{M}$-$\bar{\Gamma}$-$\bar{M}$, respectively. As was seen in the experimental plots of Fig.~\ref{fig4}(d) and (h), the surface Dirac cone spectral intensity is relatively low compared to the observed bulk states for the photon energy $h\nu$ = 75 eV. Accordingly, its contribution to the measured spin-ARPES spectra is also small. To improve the resolution of the spin information of the surface Dirac cone, the Spin-ARPES spectra in Fig.~\ref{fig5}(d) were measured with $h\nu$ = 19 eV. The crossing of the up-spin (red) and the down-spin (blue) bands are well observed around the energy of the surface Dirac point, matching well with the calculated spin texture shown in Fig.~\ref{fig5}(c). This confirms the helical nature of the spin-momentum locking around the surface Dirac point and its topological origin. The signs of measured and the calculated spin polarization is reversed for $K_{x/y} \to - K_{x/y}$, in all panels. This implies that the spin polarization is not due to the breaking of time-reversal symmetry.


Interestingly, we also observe a significant contribution of the out-of-plane component in our spin ARPES experiments and calculations for the  $\bar{K}$-$\bar{\Gamma}$-$\bar{K}$ direction,  as shown in Fig.~\ref{fig5}(e) and (f). The measurement is done with incident photons with energy $h\nu = 75$ eV. The corresponding out-of-plane spin component for the $\bar{M}$-$\bar{\Gamma}$-$\bar{M}$ direction is negligibly small. 
The scale of the in-plane and the out-of-plane spin polarization in all the panels is identical. 
%
Note that due to the presence of time reversal and inversion symmetry in CoTe$_2$, the spin polarization of the bulk states is forbidden. Even an isolated monolayer of CoTe$_2$ preserves the inversion and the time-reversal symmetries. Thus, an isolated monolayer of CoTe$_2$ will also not support spin-polarized states. 
However, in a system of finite size, the inversion symmetry is broken for the atomic layers near the surface even for bulk centrosymmetric systems. This is what allows for spin polarization of the surface states (both in-plane and out-of-plane) in a finite slab of CoTe$_2$, and other Dirac semimetals. 
Another interesting point is that the surface states near the $\bar{\Gamma}$ point primarily arise from the topological bulk band inversions, and these lead to Dirac surface states which have an in-plane Rashba-like spin momentum locking. This can be clearly seen in Fig.~\ref{fig5}(e), where the out-of-plane spin states are completely absent near the $\bar{\Gamma}$ point. 


\section{Conclusions}
\label{S7}
In summary, based on the ARPES experiments combined with detailed first principle calculations, we show that CoTe$_2$ hosts a pair of type-II Dirac nodes. The Dirac node is located along the $\Gamma-A$ axis around 0.92 eV above the Fermi energy, and they support Lorentz symmetry violating Dirac fermions. We find that in addition to the Dirac fermions, bulk CoTe$_2$ also hosts several topological band inversions which give rise to a ladder of spin-polarized surface states over a wide range of energies. The surface states corresponding to the bulk band inversions form a surface Dirac cone at the $\bar{\Gamma}$ point, which has Rashba-type in-plane spin splitting.   
Interestingly, we find that some surface states in CoTe$_2$ also support an out-of-plane spin polarization.   
Our study highlights that CoTe$_2$ supports interesting bulk and surface Dirac fermiology, which should be explored further in transport, optical, plasmonic, and optoelectronic experiments.
\vspace{1 cm}

\section{Acknowledgement}
A.C. acknowledges the Indian Institute of Technology, Kanpur, and the Science and Engineering Research Board (SERB) National Postdoctoral Fellowship (PDF/2021/000346), India for financial support. We thank Debasis Dutta and Barun Ghosh for the useful discussions. We acknowledge the Science and Engineering Research Board (SERB) and the Department of Science and Technology (DST) of the Government of India for financial support. We thank CC-IITK for providing the High-Performance Computing facility. This work has been partly performed in the framework 
of the nanoscience foundry and fine analysis (NFFA-MIUR Italy, Progetti Internazionali) facility. 
\appendix

\section{Scaled ARPES}\label{appA}
\begin{figure}
\centering
\includegraphics[width=\linewidth]{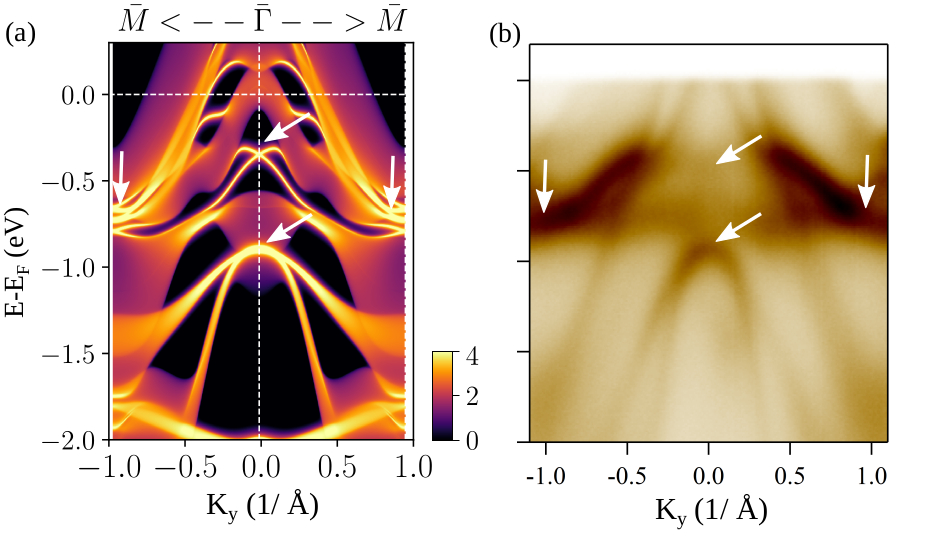}
\caption{\label{fig6} Theoretically calculated (a) and experimental (b) momentum resolved spectral function plot $\bar{M}$-$\bar{\Gamma}$-$\bar{M}$ directions. The theoretical spectral function incorporates an energy scaling factor of 0.7 to best match the experimental data.}
\end{figure}

The prominent bulk and surface states (except the Dirac crossing) of theoretically calculated spectral function and experimentally measured ARPES plots in Fig.~\ref{fig4}(c), (d) and in Fig.~\ref{fig4}(g)] have an energy difference of $\sim$500 meV. This can arise from various factors as discussed in section~\ref{S5}. For example, a similar discrepancy of energy is reported for a related compound PtSe$_2$ in Ref.~\cite{Bahramy2018U}. An energy scale factor of 1.05 and an energy offset of -0.1 eV is necessary for the PtSe$_2$ compound to correctly match the energy between theoretical and experimental ARPES results. Similarly in our calculation, an energy scaling of 0.7 can be used to best fit the experimental plot (see Fig.~\ref{fig6}). 

\section{Fermi surface anisotropy}\label{appB}
\begin{figure}[h]
\centering
\includegraphics[width=\linewidth]{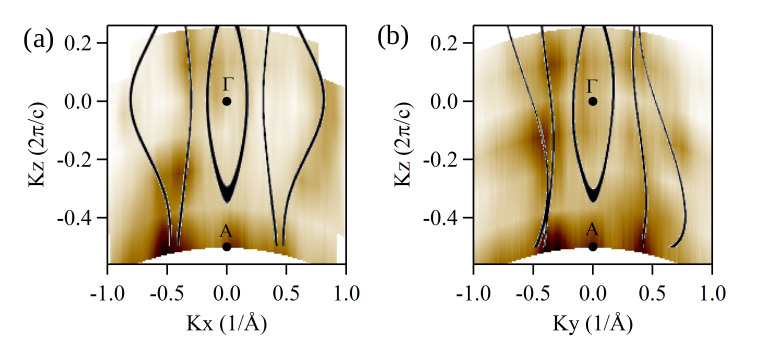}
\caption{\label{fig7} The projected experimental Fermi surface at $E=E_F$ on the (a) $K_x-K_z$ plane along where $K_x$ is along the $\Gamma$-$K$ direction, and on the (b) $K_y-K_z$ plane with $K_y$ being along the $\Gamma$-$M$ direction. The black lines are the theoretically calculated 2D energy contours. }
\end{figure}

In this section, we have compared the theoretically calculated and experimentally observed Fermi surface maps on $K_x$-$K_z$ and $K_y$-$K_z$ planes. In Fig.~\ref{fig7}, we have plotted the theoretically calculated 2D energy contours at $E=E_F$ on top of the experimental results. Here the anisotropy, as discussed in section~\ref{S4} is evident from the differences between Fig.~\ref{fig7} (a) and Fig.~\ref{fig7} (b) plots. The experimental $K_x-K_z$ and $K_y-K_z$ maps are taken with the photon energy range between 55 eV and 85 eV.

\bibliography{reference}

\end{document}